\documentclass[10pt,journal]{IEEEtranTCOM}

\usepackage[english]{babel}
\usepackage[usenames]{color}
\usepackage[cp1250]{inputenc}
\usepackage{amsfonts}
\usepackage{amsthm}
\usepackage{graphicx}
\usepackage{epsfig}
\usepackage{mathrsfs}
\usepackage{amsmath}
\usepackage{algorithm}
\usepackage{algorithmic}
\usepackage{hyperref}

\theoremstyle{plain}

\begin{document}
\newcommand{\bea}{\begin{eqnarray}}
\newcommand{\eea}{\end{eqnarray}}
\newcommand{\be}{\begin{equation}}
\newcommand{\ee}{\end{equation}}
\newcommand{\beas}{\begin{eqnarray*}}
\newcommand{\eeas}{\end{eqnarray*}}
\newcommand{\bs}{\backslash}
\newcommand{\bc}{\begin{center}}
\newcommand{\ec}{\end{center}}
\def\SC {\mathscr{C}}

\title{Time delay multi-feature correlation analysis\\ to extract subtle dependencies from EEG signals}
\author{\IEEEauthorblockN{Jarek Duda}\\
\IEEEauthorblockA{Jagiellonian University,
Golebia 24, 31-007 Krakow, Poland,
Email: \emph{dudajar@gmail.com}}}
\maketitle

\begin{abstract}
Electroencephalography (EEG) signals are resultants of extremely complex brain activity. Some details of this hidden dynamics might be accessible through e.g. joint distributions $\rho_{\Delta t}$ of signals of pairs of electrodes shifted by various time delays (lag $\Delta t$). A standard approach is monitoring a single evaluation of such joint distributions, like Pearson correlation (or mutual information), which turns out relatively uninteresting - as expected, there is usually a small peak for zero delay and nearly symmetric drop with delay. In contrast, such a complex signal might be composed of multiple types of statistical dependencies - this article proposes approach to automatically decompose and extract them. Specifically, we model such joint distributions as polynomials, estimated separately for all considered lag dependencies, then with PCA dimensionality reduction we find the dominant joint density distortion directions $f_v$. This way we get a few lag dependent features $a_i(\Delta t)$ describing separate dominating statistical dependencies of known contributions: $\rho_{\Delta t}(y,z)\approx \sum_{i=1}^r a_i(\Delta t)\, f_{v_i}(y,z)$. Such features complement Pearson correlation, extracting hidden more complex behavior, e.g. with asymmetry which might be related with direction of information transfer, extrema suggesting characteristic delays, or oscillatory behavior suggesting some periodicity. There is also discussed extension of Granger causality to such multi-feature joint density analysis, suggesting e.g. two separate causality waves. While this early article is initial fundamental research, in future it might help e.g. with understanding of cortex hidden dynamics, diagnosis of pathologies like epilepsy, determination of precise electrode position, or building brain-computer interface. 
\end{abstract}
\textbf{Keywords:}  electroencephalography (EEG), multi-feature correlation analysis, time series analysis, time shift, signal processing, Granger causality, principal component analysis (PCA), hierarchical correlation reconstruction (HCR)
\section{Introduction}
Electroencephalography (EEG) allows for noninvasive, high temporal resolution,  and relatively inexpensive  monitoring of cerebral cortex activity, making it a very popular technique for basic research and applications like medical diagnosis (e.g. epilepsy), brain-computer interface (BCI), economy/marketing research. Its signals average extremely complex hidden dynamics into a relatively small number of time series - e.g. 32 electrodes sampled 500 Hz in the used example analysis (data source: \footnote{\url{https://www.kaggle.com/competitions/grasp-and-lift-eeg-detection/data}} \cite{source}).

\begin{figure}[t!]
    \centering
        \includegraphics[width=9.5cm]{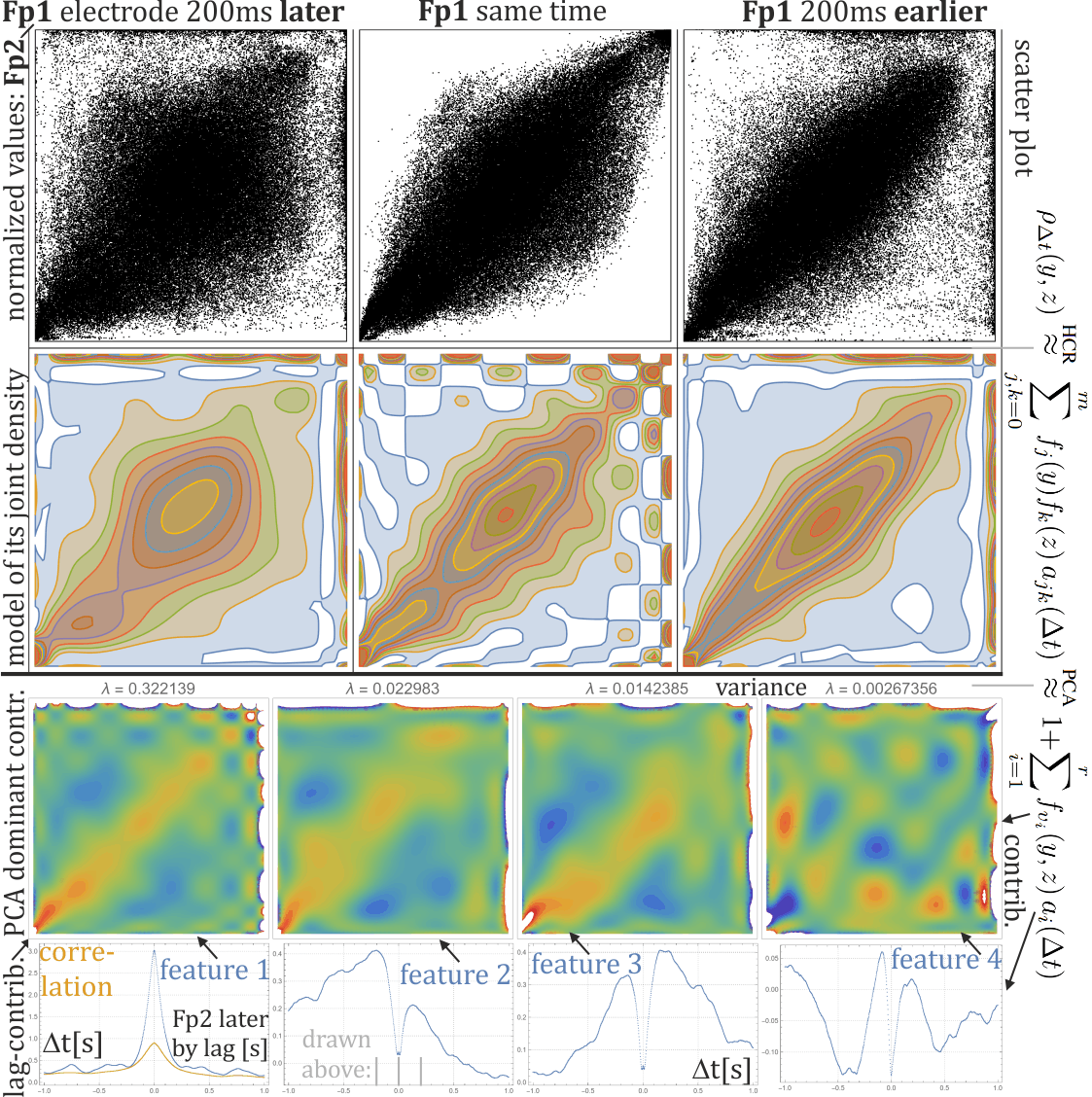}
        \caption{Proposed analysis on example of Fp1-Fp2 pair of electrodes. The signals are first independently normalized to nearly uniform on [0,1] with Gaussian CDF: $y_t = \textrm{CDF}_{N(0,1)}((x_t - E[X])/\sqrt{\textrm{var}(X)})$. The top diagram shows scatterplots for $\approx 140 000$ pairs of normalized values in the same time, and shifted by 200ms in both time directions. Pearson correlation coefficient, plotted orange in the bottom, as expected weakens with such delay in nearly symmetric way. In contrast,  we can see the behavior is much more complex, e.g. clearly lag asymmetric - requiring more sophisticated analysis. The used analysis models the joint distributions for each lag as a polynomial (second row, HCR). Then for each electrode pair the basis is reduced to PCA optimized: maximizing variance over the considered delays. In third row there are shown such $r=4$ dominant contributions to joint distribution (linear combinations of the original ones).  The final features are shown in the bottom - they are lag/delay dependencies of the dominating contributions to joint distribution. Further are shown in Fig. \ref{multi}, \ref{multifeature}, often containing asymmetry suggesting directionality of information transfer, extrema suggesting characteristic delays, or oscillations suggesting periodicity.}
       \label{intr}
\end{figure}

This article focuses on statistical dependencies of such time series for pairs of electrodes shifted by various delays (up to 1 second in the examples here): different electrodes for cross-correlation, the same for autocorrelation analysis. Standard approaches usually monitor correlation or mutual information~\cite{del1,del2,del3,del4,del5}. In contrast, extreme complexity of the represented hidden activity suggests to search for more subtle multiple separate, complementing dependencies - we automatically find here, and they turn out to have much more complex behavior than the standard (Pearson) correlation coefficient.

The proposed approach is based on Hierarchical Correlation Reconstruction (HCR)~\cite{hcr} family of methods, which decompose dependencies into multiple mixed moments chosen to represent joint distribution as a linear combination. Similar multi-feature correlation analysis was previously performed for autocorrelation~\cite{cor1} and cross-correlations~\cite{cor2}, here combined into time delay multi-feature correlation analysis: between two (or more) series shifted in time.

\begin{figure*}[t!]
    \centering
        \includegraphics[width=19cm]{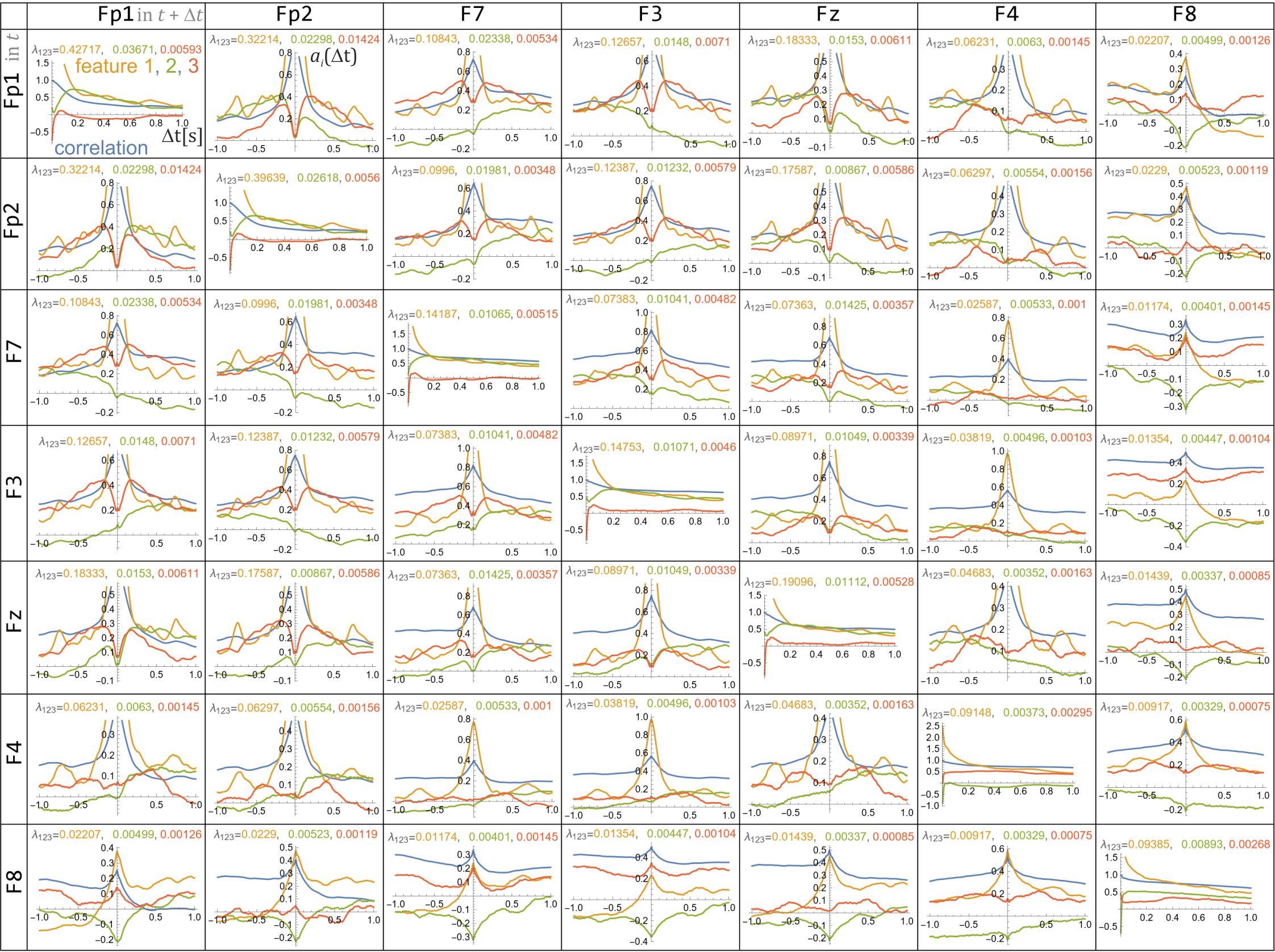}
        \caption{We approximate joint density lag dependance as $\rho_{\Delta t}(y,z)\approx \sum_{i=1}^r a_i(\Delta t)\, f_{v_i}(y,z)$, this Figure contains  $a_{i} (\Delta t) \equiv a_{v_i} (\Delta t)$ for all pairs for the first 7 electrodes and $r=3$ features (orange, green, red), also Pearson correlation for comparison (blue). In diagonal there are autocorrelations (only one sign of time shift $\Delta t \in (0,1]$s), outside diagonal there are cross-correlations ($\Delta t \in (-1,1]$s). Analogously for all 32 electrodes is presented in Figure \ref{multifeature}. Corresponding PCA eigenvalues $\lambda$ are written above, they are the maximized variances of considered $\{a_v(\Delta t)\}$. The first feature (orange) is usually similar to Pearson correlation (blue), but having much stronger localization in $\Delta t =0$. More interesting seem the novel remaining two features: green one is often asymmetric, what might suggest information transfer direction. The red one often has maximal dependance for $\Delta t\sim 200$ms delay.
        }
       \label{multi}
\end{figure*}

\begin{figure*}[t!]
    \centering
        \includegraphics[width=19cm]{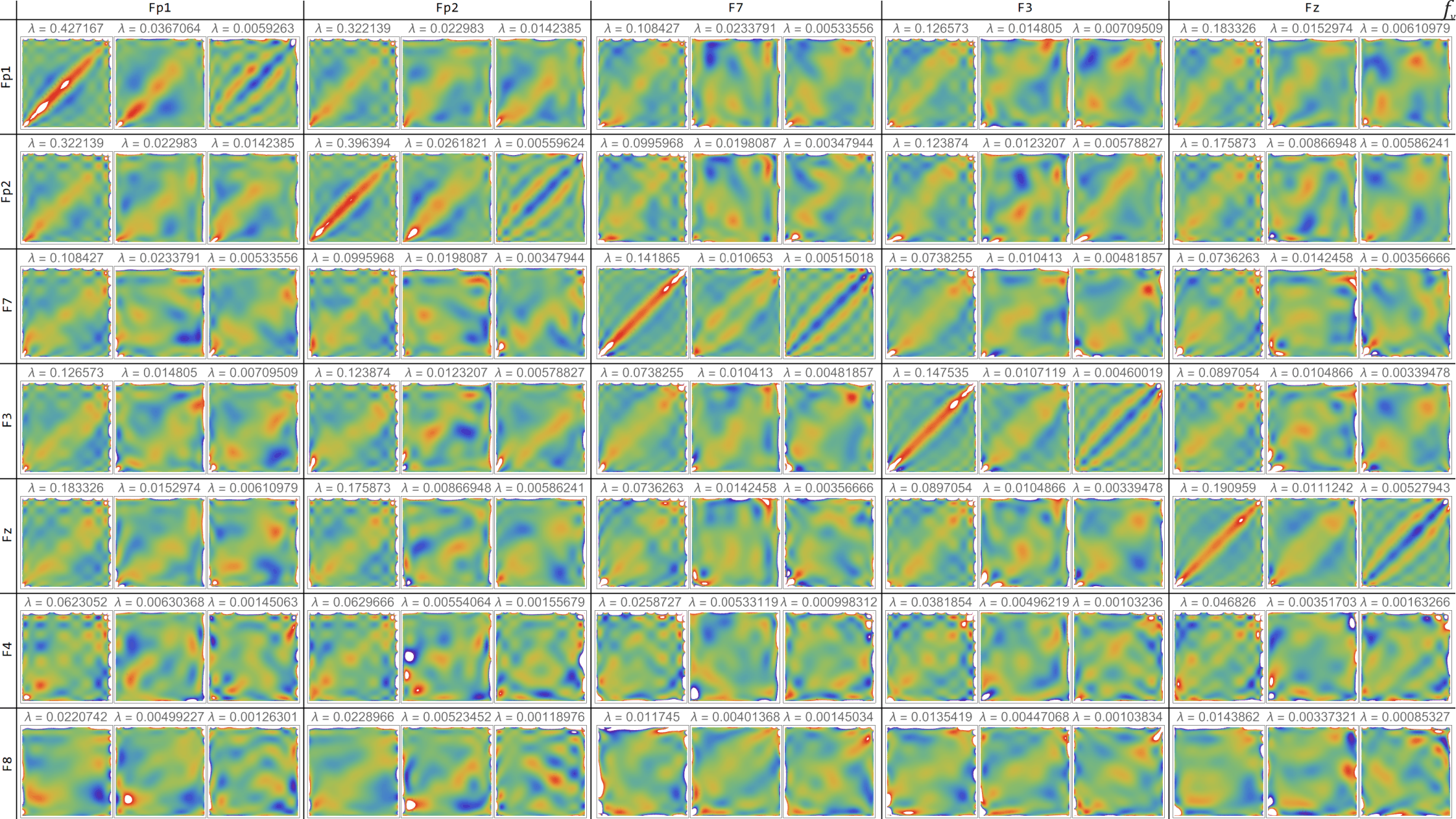}
        \caption{PCA optimized dominant contributions $f_v$ to joint distribution of normalized variables for some pairs of electrodes (red: positive, blue: negative), approximating joint density as $\rho_{\Delta t}(y,z)\approx \sum_{i=1}^r a_i(\Delta t)\, f_{v_i}(y,z)$. Their lag dependent contributions $a_{v_i}$ are the discussed features presented in Fig. \ref{multi}, \ref{multifeature}. These $f_v$ contributions allow for interpretation of $a_v(\Delta t)$ features, and might themselves contain some valuable statistical information about the hidden cortex dynamics, e.g. to identify the functional structure attached to the electrode. For autocorrelations the above contributions are close to diagonal, with $a_v$ features describing details of blurring of this diagonal with growing lag. For cross-correlations the diagonal is often also emphasized, but generally the behavior is much more complex. We can also observe asymmetry between bottom-left and top-right corners, describing cooccurrences of the lowest and the highest extreme events.}
       \label{contr}
\end{figure*}

Example of the discussed analysis is shown in Fig. \ref{intr} for Fp1-Fp2 pair of electrodes. While scatterplots show clear delay asymmetry, it is neglected in Pearson correlation coefficient (orange plot at the bottom), but clearly seen in the proposed features (blue plots). Specifically, like in copula theory~\cite{copula}, it is convenient to first normalize variables to nearly uniform  distributions on [0,1] by applying cumulative distribution function ($y_t=\textrm{CDF}(x_t)$), here of Gaussian distribution. Then considering pairs of such normalized variables, if independent they would be from nearly uniform distribution on $[0,1]^2$. As they are dependent, in HCR we model their joint density as a linear combination in orthonormal (product) basis for some chosen set (here two variables, $m=10$, $r=3$ or $4$, $a_i \equiv a_{v_i}$):
\be \rho(y,z)\stackrel{\text{HCR}}{\approx} \sum_{j,k=0}^m a_{jk} f_j(y) f_k(z)
   \stackrel{\text{PCA}}{\approx}1+ \sum_{i=1}^r  a_i\, f_{v_i}(y,z)\ee

\noindent for some 1D $f_j$ orthonormal basis ($\int_0^1 f_j(y) f_k(y) dy =\delta_{jk}$), here Legendre polynomials. It starts with $f_0=1$, hence $a_{00}=1$ term corresponds to normalization, $a_{j0}$ and $a_{0k}$ terms describe marginal distributions. Term $a_{11}$ is close to Pearson correlation coefficient, $a_{jk}$ mixed moment describes dependence between $j$-th moment of the first variable, and $k$-th moment of the second variable. Here there was used arbitrarily chosen $m=10$, what means 100 mixed moments (plus subtracted in analysis 20 marginal moments) - while it seems a lot, we can still see artifacts in the central density in Fig. \ref{intr}, suggesting to consider even larger $m$, or starting with a different basis of functions.

Now shifting one of two considered time series (electrodes) by $\Delta t$ lag, the coefficients become dependent on this lag: $a_{kl}\equiv a_{kl}(\Delta t)$. As we are talking about a hundred of lag dependent coefficients, we can apply a dimensionality reduction technique like PCA to choose a few dominant features $a_v(\Delta t)$, being linear combinations of the original ones $a_{kl}(\Delta t)$, chosen to maximize variance over the considered set of lags, here of 1001 lags: from -1s to +1s with 500Hz sampling, independently for each pair of electrodes. 

As we will see in the figures, the first such feature is usually similar to standard (Pearson) correlation - nearly symmetric with peak for zero delay, however, it is more localized as this statistical dependency is more diagonal than standard correlation. Especially for autocorrelation for which zero lag would mean joint density being a perfect diagonal (here we start with the smallest nonzero lag). More interesting are the further features - often having asymmetry, local extrema, oscillations, etc. - which might correspond to properties of hidden cortex dynamics.

The second version of this article adds simple extension to Granger's causality - by just adding prediction from its past to the later sequence in the considered joint distribution. Such multi-feature causality e.g. distinguishes separate causality waves, rather merged in standard single  causality evaluations.

The current early version of article presents basic methodology and examples (calculated in Wolfram Mathematica), to be extended in the future - for example with further techniques e.g. from \cite{cor1,cor2}, like inclusion of nonstationarity analysis. Natural development directions are applications: from basic research improving understanding of cortex activity through the discussed features, to e.g. medical diagnosis or brain-computer interface applications.


\section{Delay time multi-feature covariance analysis}\label{sec2}
This main section introduces to the proposed methodology, in the next section asymmetrized by adding prediction from its past to normalization of the later sequence (Granger causality).
\subsection{Data and normalization}
All presented example analysis was made using EEG data from a single file in the mentioned Kaggle competition "Grasp-and-Lift EEG Detection": 32 electrodes $\times$ 140424 times with 500Hz sampling ($\approx 280$ seconds). The purpose of this early version of article is introduction to the methodology, to provide basic intuitions from examples. In future there should be analyzed and  compared multiple samples to investigate universality. 

As in copula theory~\cite{copula}, in the discussed HCR methodology it is convenient (not necessary) to normalize all data (each time series individually) to nearly uniform distributions $y_t=\textrm{CDF}(x_t)$. There were tested various distributions and Gaussian turns out nearly optimal, hence it was used (also for simplicity). Specifically, for each time series $\{x_t\}$ there was independently calculated mean and variance, then we further work on normalized $\{y_t\}$:   $y_t = \textrm{CDF}_{N(0,1)}((x_t - E[X])/\sqrt{\textrm{var}(X)})$ using CDF of normalized Gaussian distribution ($\mu=0,\sigma=1$).

\begin{figure*}[t!]
    \centering
        \includegraphics[width=19cm]{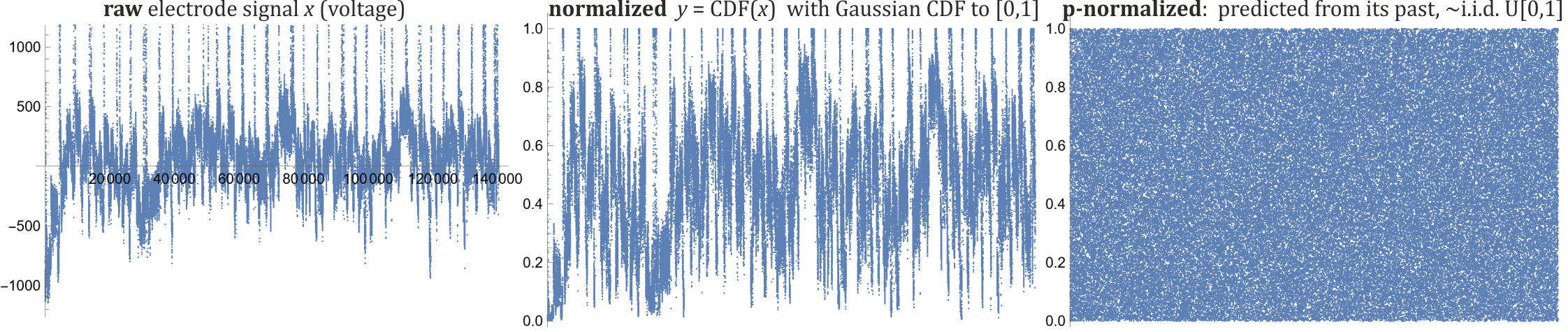}
        \caption{The normalization process example - the left hand side plot shows the original time series (from Fp1 electrode). The middle plot shows basic \textbf{normalization} used in Section \ref{sec2}: subtraction of mean, division by standard deviation, and applying CDF of Gaussian distribution: $y_t = \textrm{CDF}_{N(0,1)}((x_t - E[X])/\sqrt{\textrm{var}(X)})$, making $\{y_t\}_t$ set from nearly uniform $U[0,1]$ distribution. The right hand side plot shows much stronger predicted \textbf{p-normalization} used for causality analysis in Section \ref{sec3}: using the entire past of this time series to predict separate $\textrm{CDF}_t$ for each time and $z_t=\textrm{CDF}_t (x_t)$ normalization, making $(z_t)_t$ series very close to i.i.d. $U[0,1]$. The current prediction was order 10 ARMA, then order 1 ARCH, then adaptive Student's t distribution with optimized $\nu\sim 10$ shape parameter. For causality evaluation we analyze distortion from uniform joint distribution between some earlier normalized series (reason) and later p-normalized different series. }
       \label{norm}
\end{figure*}

However, as we can see in the middle plot of Fig. \ref{norm}, such normalized time series is rather nonstationary. From one side there is an open question if it should be removed for correlation analysis, what can be done using predicted CDF with time dependent parameters: $y_t=\textrm{CDF}_t(x_t)$ - done as p-normalization in the right hand plot of Fig. \ref{norm} for Granger causality in the next Section. From the other side, nonstationarity hides crucial information about activity - worth being analyzed, what can be combined with the discussed HCR by just replacing averages in estimation with exponential moving averages to estimate time evolution e.g. of the discussed features, as for example in \cite{cor1}.
\subsection{Hierarchical correlation analysis (HCR)}
The HCR philosophy~\cite{hcr} models (joint) densities as linear combinations $\rho(y)\approx \sum_j a_j f_j(y)$, allowing for inexpensive to estimate high parameter description, decomposition of statistical dependencies into mixed moments - intuitively like Taylor expansion of multivariate function.

For estimation it is convenient to use orthonormal family of functions: $\int_0^1 f_j(y) f_k(y) dy =\delta_{jk}$, usually polynomials, alternatively could be e.g. trigonometric especially for periodic data. Here we use othonormal (Legendre) polynomials - $f_0,f_1,f_2,f_3,f_4$ are correspondingly:
$$ 1,\sqrt{3}(2x-1), \sqrt{5}(6x^2-6x+1),$$
$$\sqrt{7}(20x^3-30x^2+12x-1),\  (70x^4-140x^3+90x^2-20x+1).$$
In multiple dimensions a natural approach is using product basis, e.g. here starting with $f_{jk}(y,z)=f_j(y) f_k(z)$ for $j,k=0,\ldots,m$ and arbitrarily chosen $m=10$. Later the final $r$ features are linear combinations, in basis optimized with PCA.

For orthonormal basis, MSE estimation is simple, e.g. used 2D for $P=\{(y,z)\}$ dataset - pairs of (normalized) values:
\be \rho(y,z)\approx \sum_{j,k=0}^m a_{jk} f_j(y)f_k(z)\quad\textrm{for}\quad a_{jk} = \underset{(y,z)\in P}{\textrm{mean}}\, f_j(y)f_k(z) \label{est}\ee
allowing to estimate $a_{jk}$ independently, also e.g. their evolution by replacing average with exponential moving average. Estimation of such single $\approx 100$ parameter joint distribution takes a few milliseconds on a notebook.

\subsection{PCA dimensionality reduction}
For the discussed application we estimate parameters (\ref{est}) separately for each introduced delay/lag $\Delta t$, getting $a_{kl}(\Delta t)$ three dimensional tensor - here for each electrode pair: $11 \times 11 \times 1001$ for $m=10$ and 1 second maximal delay at 500Hz sampling.

We need to reduce it for practical applications - here using PCA. For this purpose, this tensor was first treated as $121 \times 1001$ matrix, then its $121 \times 121$ covariance matrix $C$ was calculated, and $r=3$ or 4 its dominant eigenvectors $Cv=\lambda v$ were calculated.

Then we define $f_v = v\cdot (f_{jk}:j,k=0,\ldots,m)$ as the new optimized basis of dominant joint distribution contributions, with transformed coefficients: $a_v(\Delta t) = v\cdot (a_{jk}(\Delta t):j,k=0,\ldots,m)$. The corresponding eigenvalue $\lambda$ is variance of such lag dependent features $\{a_v(\Delta t)\}$. Some of $f_v$ are shown in Figure \ref{contr} - they are dominant contributions to joint distribution with lag dependence given by $a_v(\Delta t)$.

Additionally, we can remove contribution of marginal distributions before PCA, what was applied but is nearly negligible here (more important for Granger causality in the next Section), using the below modified coefficients:
\be   \tilde{a}_{jk}=a_{jk}-a_{j0}\,a_{0k}\qquad \textrm{for}\quad k,j>0 \label{marg} \ee

\begin{figure*}[t!]
    \centering
        \includegraphics[width=19cm]{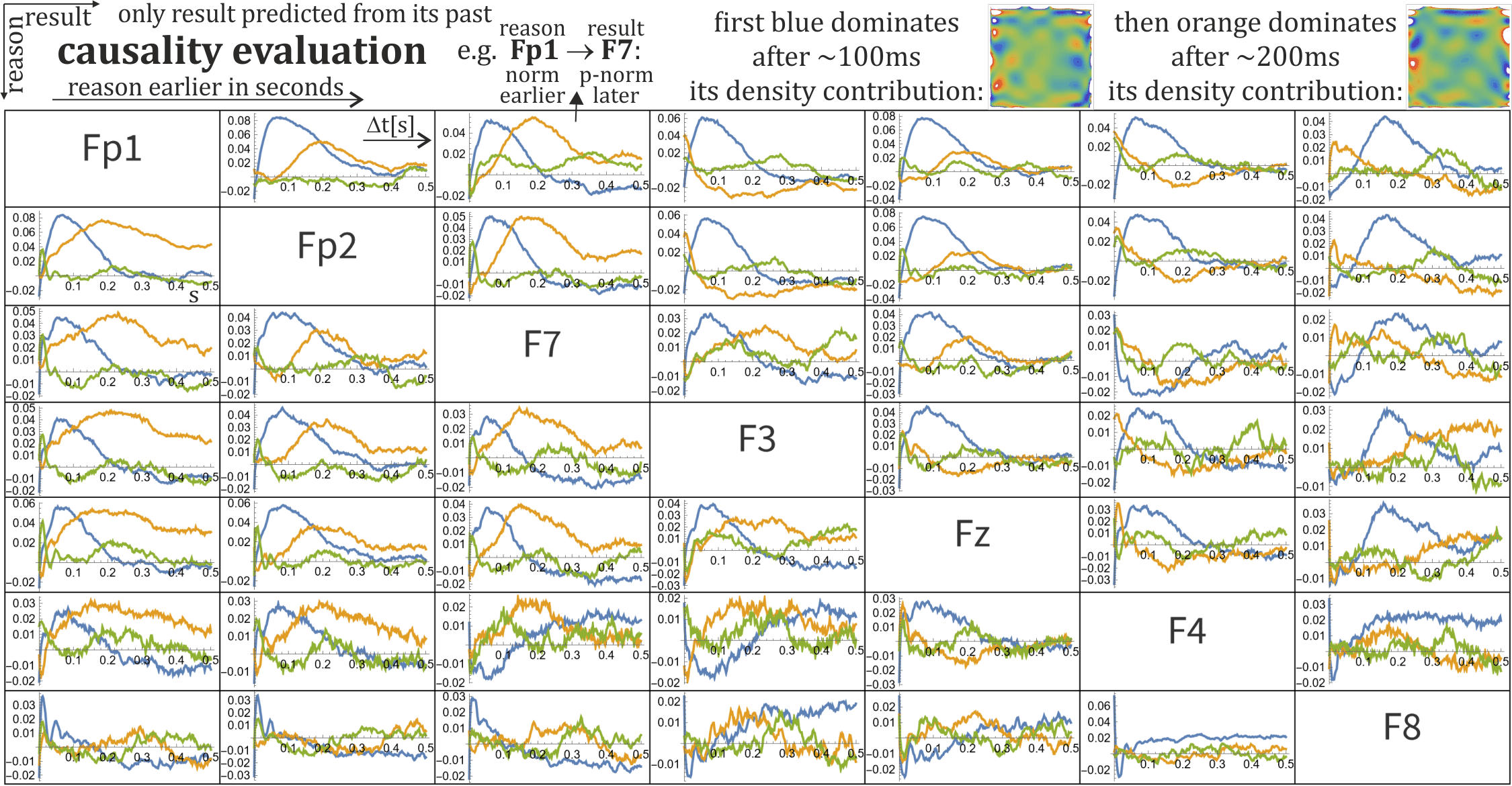}
        \caption{Full causality evaluation for the first 7 electrodes (for all summarized in Fig. \ref{causeval}). As the reason (row) we use basic normalized earlier signal from one electrode ($A$), as the result (column) we use predicted from its past p-normalized later signal from a different electrode ($B$). There are plotted the first three PCA optimized multi-feature cross-covariance analysis features - the dominant distortions of joint density for such pairs of signals. For some like emphasized example Fp1 $\to$ F1 we can see kind of two separate (shown orthogonal joint density distortions) information transfer waves: $\approx $ 100ms blue wave, and then later $\approx $ 200ms orange wave - distinguished in the discussed multi-feature approach, would probably merge into one if using e.g. mutual information instead. We can also see asymmetry, e.g. Fp1$\to$F3 has much weaker this second orange wave than the opposite (sign is negligible).   }
       \label{causplot}
\end{figure*}

\begin{figure}[t!]
    \centering
        \includegraphics[width=8.5cm]{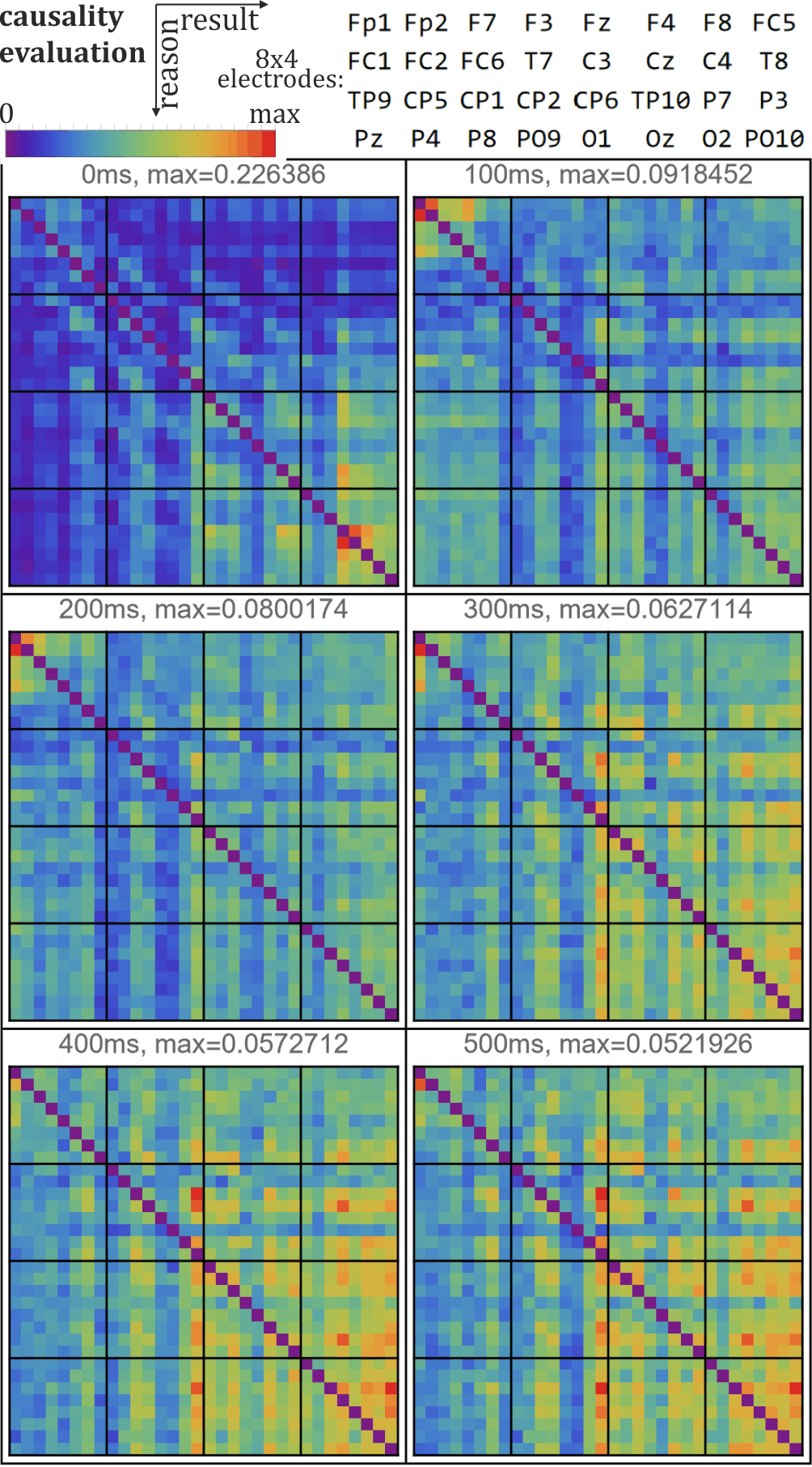}
        \caption{Proposed causality evaluation for various delays. Time series of predicted electrode ($B$, horizontal, result) was p-normalized (with prediction). Time series of second electrode, earlier the written delay, was just normalized with Gaussian distribution ($A$, vertical, reason). There are shown causality evaluations as square root of sum of squared discussed features $\sqrt{\sum_v (a_v)^2}$, describing distortion from uniform joint distribution. While these diagrams use the same colors, they refer to different scales - up to the written "max". Very large evaluation for zero delay is characteristic here for visual cortex, and suggests some earlier common (visual) cause. For 100-200ms delays the strongest causality is evaluated for  frontal cortex, for higher delays we can see strong asymmetry - especially toward T8 electrode.
        }
       \label{causeval}
\end{figure}

\begin{figure*}[t!]
    \centering
        \includegraphics[width=18cm]{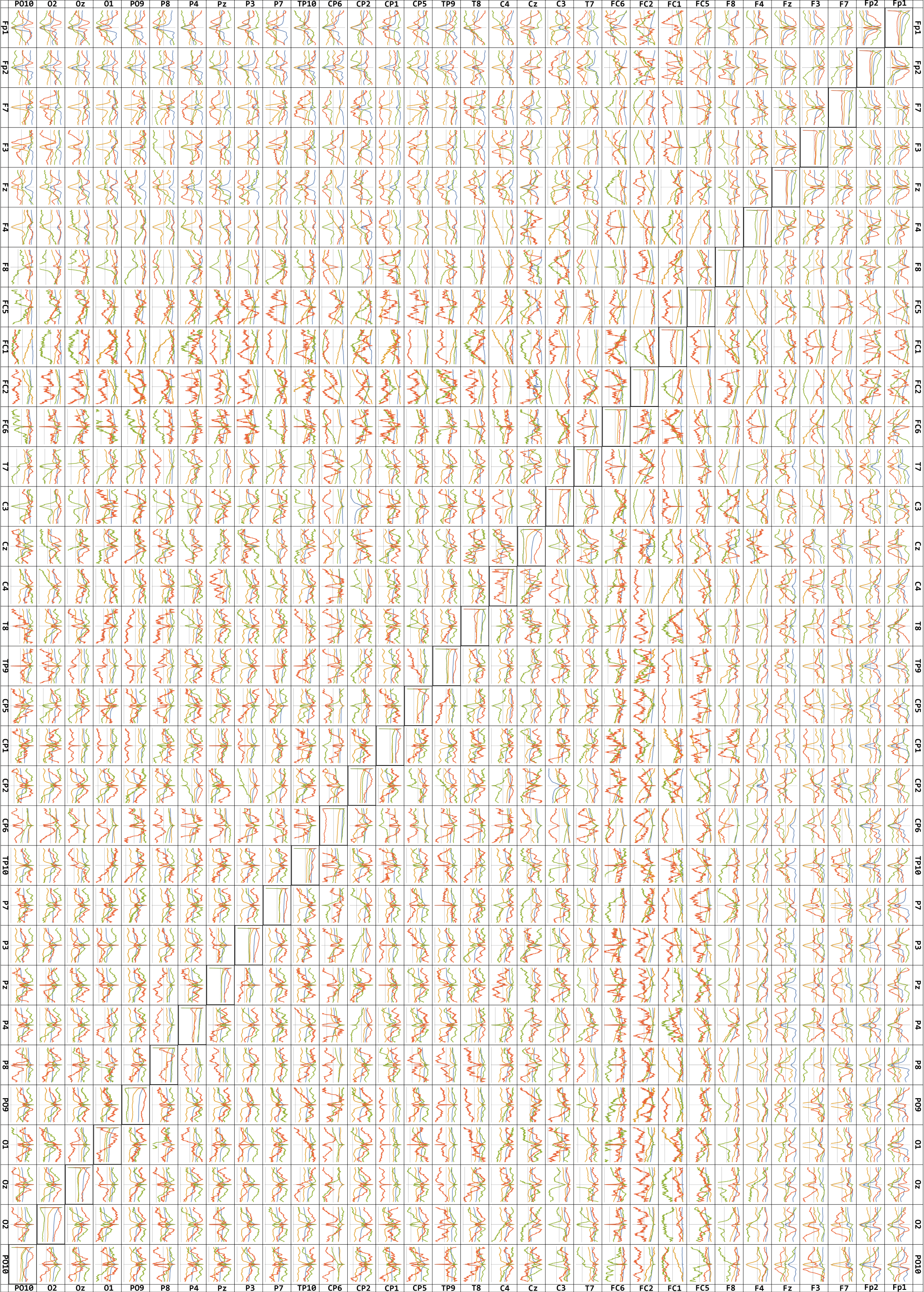}
        \caption{As in Figure \ref{multi}, this time for all 32 electrodes, for better readability plots were normalized to $a_v(\Delta t)/\max_{\tau} |a_v(\tau)|$. }
       \label{multifeature}
\end{figure*}

\section{Density-based multi-feature Granger causality} \label{sec3}
In very popular Granger causality~\cite{granger}, \emph{if $A$ signal "\textbf{Granger-causes}" a $B$ signal, then past values of $A$  should contain information that helps predict $B$ above and beyond the information contained in past values of $B$ alone.} Usually it is done with very simple linear regression value prediction like in ARMA models. 

Value prediction alone cannot see causality as e.g. influence on variance and higher moments. To observe such more subtle dependencies, we need to switch from working on values, to probability distributions. There are popular approaches working on relatively simple distributions like multivariate Gaussian, however, especially in EEG data we can see the joint distributions are much more complex - suggesting to use agnostic large number of parameters models, like the discussed fitting polynomial to joint distributions. There are ways to directly predict such simple models of distribution as polynomial (e.g. \cite{cred}), however, for simplicity let us now just modify the above approach here, maybe to use a different one in some future.

Specifically, looking at the above description of Granger causality, from the predicted series ($B$, later result) we need first to remove information which can be concluded from its past. For value prediction we just subtract the prediction, e.g. using linear regression from moving window in ARMA models, then can try to predict the difference (called residue) for Granger causality using the second sequence ($A$, earlier reason).

To analogously "subtract" predicted from previous values entire probability distributions, e.g. used here Student's t-distributions centered in the predicted value, a natural approach is predicting this probability distribution as $\textrm{CDF}_t$ in time $t$, and normalizing $z_t=\textrm{CDF}_t(x_t)$ to nearly i.i.d. uniform $U[0,1]$. We will refer to it as \textbf{p-normalization}, with example in Fig. \ref{norm}. 

Having such p-normalized sequence for the $B$ signal with removed information about its past, as Granger causality we can search for its statistical dependencies from earlier $A$ signal - here with basic Gaussian normalization. We could use various approaches to evaluate the dependence e.g. conditional entropy, mutual information, Kullback-Leibler~\cite{transfer}, however, all of them are some features of joint distribution - instead of extracting such feature, we can directly describe (approximated but) complete joint distributions as in the discussed multi-feature philosophy: through dominant (PCA optimized) distortions of $\rho=1$ starting joint distribution. 

Such multi-feature approach allows to decompose Granger causality into complementary, literally orthogonal contributions to joint distributions between earlier $B$ sequence, and later $A$ sequence having "subtracted" prediction from its past. As we can see in Figure \ref{causplot}, such orthogonal features can suggest e.g. two causality waves (blue $\approx 100$ms then orange $\approx 200$ms), with concrete interpretations as shown join distribution contributions. Such two waves would likely be merged if extracting single evaluation from joint distribution like mutual information.

While Figure \ref{causplot} plots allow for detailed analysis of pairs of electrodes as $\rho_{\Delta t}(y,z)\approx 1+\sum_v a_v(\Delta t)\, f_v(y,z)$, a simple way to combine them into a single causality evaluation (alternative e.g. for mutual information) is $\sqrt{\sum_v (a_v)^2}$, thanks to orthonormality of $f_v$ approximately mean deviation of $\rho\approx 1$ uniform joint density. Such evaluation for all electrode pairs and various delays is shown in Fig. \ref{causeval}. We can e.g. see optical cortex having vary large such causality evaluation for 0 delay - suggesting a common cause like visual stimulus. For $\sim 50-200$ ms delays the strongest evaluation is for the frontal cortex. The rest of cortex dominates for higher delays - often in clearly asymmetric way.

For the used p-normalization, there was applied the following procedure optimized for log-likelihood, which leads to nearly perfect i.i.d. U(0,1) normalized series, but still can be improved in the future. Specifically, there was first subtracted prediction from linear regression using 10 previous values (ARMA). Then such residues were divided by standard deviations predicted by order 1 ARCH model. For such sequence there was applied recent adaptive Student't distribution~\cite{student} estimating evolving scale parameter from exponential moving averages, for fixed but individually (log-likelihood) optimized $\nu\sim 10$ shape parameter, finally used for $z_t=\textrm{CDF}_t (x_t)$ p-normalization.

\section{Conclusions and further work}
The article proposes time delay multi-feature correlation analysis - decomposing statistical dependencies into a few dominant lag dependent features, to automatically optimize and extract multiple types of subtle statistical dependencies, causality evaluations e.g. for EEG electrodes, but also for other types of time series.

This is early version of methodology and article, introducing this looking promising novel approach, requiring further work - starting e.g. with comparison of such analysis for various activities and persons, trying to understand hidden mechanisms leading to these  dominant statistical dependencies.

Some possible future research directions:
\begin{itemize}
  \item Practical applications, like medical diagnosis, detection of pathological activity e.g. seizure, determination of precise electrode position, various brain-computer interface applications.
  \item Better understanding based on such found features - of information transfer, cortex activity, their interpretations and properties - like various asymmetries, extrema for characteristic delays, oscillations for some periodicity, suggested two causality waves with asymmetries, etc. 
  \item The example analysis was performed on a single time series for 32 electrodes - bringing a big question of universality: dependence of electrode positions, person, conditions, anatomical changes like cerebral vascular accident or corpus callosotomy procedure disconnecting hemispheres. 
  \item The suggested multi-feature Granger causality allows e.g. to split causality into separate waves, suggesting further research of propagation of such waves, e.g. with higher spatial resolution (more electrodes).  
  \item While we have focused on EEG electrodes, the proposed approach is very general, allowing e.g. to combine with other time series data like  fMRI, NIRS, MEG, simultaneous electrocardiogram, etc., or use it for completely different types of data (especially multi-feature Granger causality) - suggesting search and exploration of potential applications.
  \item While it is natural to focus on lag dependence $a_v (\Delta t)$, also the found contributions $f_v$ might turn out carrying valuable information, e.g. helping in identification of the attached functional cortex region.
  \item The signals are very nonstationary, what might be worth including in normalization (e.g. as in \cite{adapt}, p-normalization for Granger causality), also perform nonstationarity analysis (e.g. as in \cite{cor1}) for example to correlate evolution of such features with stimuli/actions for brain-compute interface applications.
  \item Having discussed time dependance (here of delay), we can perform its Fourier analysis - which might provide characteristic spectra, phase shifts, it generally might be different for various discussed features - allowing to find some hidden more subtle periodic processes.
  \item While we have focused on two dimensional joint distributions, it can be easily expanded to work with 3 or more e.g. electrodes and/or delays, by starting with higher dimensional product bases.
\end{itemize}

\bibliographystyle{IEEEtran}
\bibliography{cites}

\begin{thebibliography}{10}
\providecommand{\url}[1]{#1}
\csname url@samestyle\endcsname
\providecommand{\newblock}{\relax}
\providecommand{\bibinfo}[2]{#2}
\providecommand{\BIBentrySTDinterwordspacing}{\spaceskip=0pt\relax}
\providecommand{\BIBentryALTinterwordstretchfactor}{4}
\providecommand{\BIBentryALTinterwordspacing}{\spaceskip=\fontdimen2\font plus
\BIBentryALTinterwordstretchfactor\fontdimen3\font minus
  \fontdimen4\font\relax}
\providecommand{\BIBforeignlanguage}[2]{{%
\expandafter\ifx\csname l@#1\endcsname\relax
\typeout{** WARNING: IEEEtran.bst: No hyphenation pattern has been}%
\typeout{** loaded for the language `#1'. Using the pattern for}%
\typeout{** the default language instead.}%
\else
\language=\csname l@#1\endcsname
\fi
#2}}
\providecommand{\BIBdecl}{\relax}
\BIBdecl

\bibitem{source}
M.~D. Luciw, E.~Jarocka, and B.~B. Edin, ``Multi-channel eeg recordings during
  3,936 grasp and lift trials with varying weight and friction,''
  \emph{Scientific data}, vol.~1, no.~1, pp. 1--11, 2014.

\bibitem{del1}
J.~Bonita, L.~Ambolode, B.~Rosenberg, C.~Cellucci, T.~Watanabe, P.~Rapp, and
  A.~Albano, ``Time domain measures of inter-channel eeg correlations: a
  comparison of linear, nonparametric and nonlinear measures,'' \emph{Cognitive
  neurodynamics}, vol.~8, pp. 1--15, 2014.

\bibitem{del2}
P.~Boeijinga and F.~L. da~Silva, ``A new method to estimate time delays between
  eeg signals applied to beta activity of the olfactory cortical areas,''
  \emph{Electroencephalography and clinical neurophysiology}, vol.~73, no.~3,
  pp. 198--205, 1989.

\bibitem{del3}
T.~M. Ellmore, K.~Ng, and C.~P. Reichert, ``Early and late components of eeg
  delay activity correlate differently with scene working memory performance,''
  \emph{PloS one}, vol.~12, no.~10, p. e0186072, 2017.

\bibitem{del4}
A.~Lin, K.~K. Liu, R.~P. Bartsch, and P.~C. Ivanov, ``Delay-correlation
  landscape reveals characteristic time delays of brain rhythms and heart
  interactions,'' \emph{Philosophical Transactions of the Royal Society A:
  Mathematical, Physical and Engineering Sciences}, vol. 374, no. 2067, p.
  20150182, 2016.

\bibitem{del5}
K.~Kim, S.-H. Lim, J.~Lee, W.-S. Kang, C.~Moon, and J.-W. Choi, ``Joint maximum
  likelihood time delay estimation of unknown event-related potential signals
  for eeg sensor signal quality enhancement,'' \emph{Sensors}, vol.~16, no.~6,
  p. 891, 2016.

\bibitem{hcr}
J.~Duda, ``Hierarchical correlation reconstruction with missing data, for
  example for biology-inspired neuron,'' \emph{arXiv preprint
  arXiv:1804.06218}, 2018.

\bibitem{cor1}
J.~Duda and G.~Bhatta, ``Gamma-ray blazar variability: New statistical methods
  of time-flux distributions,'' \emph{Monthly Notices of the Royal Astronomical
  Society}, 2021.

\bibitem{cor2}
J.~Duda, H.~Gurgul, and R.~Syrek, ``Multi-feature evaluation of financial
  contagion,'' \emph{Central European Journal of Operations Research}, pp.
  1--28, 2021.

\bibitem{copula}
F.~Durante and C.~Sempi, ``Copula theory: an introduction,'' in \emph{Copula
  theory and its applications}.\hskip 1em plus 0.5em minus 0.4em\relax
  Springer, 2010, pp. 3--31.

\bibitem{granger}
C.~W. Granger, ``Testing for causality: A personal viewpoint,'' \emph{Journal
  of Economic Dynamics and control}, vol.~2, pp. 329--352, 1980.

\bibitem{cred}
J.~Duda and A.~Szulc, ``Credibility evaluation of income data with hierarchical
  correlation reconstruction,'' \emph{arXiv preprint arXiv:1812.08040}, 2018.

\bibitem{transfer}
T.~Schreiber, ``Measuring information transfer,'' \emph{Physical review
  letters}, vol.~85, no.~2, p. 461, 2000.

\bibitem{student}
J.~Duda, ``Adaptive student's t-distribution with method of moments moving
  estimator for nonstationary time series,'' \emph{arXiv preprint
  arXiv:2304.03069}, 2023.

\bibitem{adapt}
------, ``Adaptive exponential power distribution with moving estimator for
  nonstationary time series,'' \emph{arXiv preprint arXiv:2003.02149}, 2020.

\end{thebibliography}
\end{document}